\documentclass[twocolumn,citeautoscript,showpacs,preprintnumbers,amsmath,amssymb,prb,floatfix,footinbib]{revtex4}
\usepackage{graphicx}
\usepackage{dcolumn}
\usepackage{bm}
\usepackage{times}
\begin{document}
\title{Antiferromagnetic ordering in the absence of a structural distortion in Ba(Fe$_{1-x}$Mn$_{x}$)$_{2}$As$_{2}$}
\author{M. G. Kim,$^1$ A. Kreyssig,$^1$ A. Thaler,$^1$ D. K. Pratt,$^1$ W. Tian,$^1$ J. L. Zarestky,$^1$ M. A. Green,$^{2,3}$ S. L. Bud$'$ko,$^1$ P. C. Canfield,$^1$ R. J. McQueeney$^1$ and A. I. Goldman$^1$}
\affiliation{\\$^1$Ames Laboratory, U.S. DOE and Department of
Physics and Astronomy, Iowa State University, Ames, IA 50011, USA}
\affiliation{\\$^2$NIST Center for Neutron Research, National
Institute of Standards and Technology, Gaithersburg, MD 20899, USA}
\affiliation{\\$^3$Department of Materials Science and Engineering,
University of Maryland, College Park, Maryland 20742, USA}
\date{\today}

\begin{abstract}
Neutron and x-ray diffraction studies of
Ba(Fe$_{1-x}$Mn$_x$)$_2$As$_2$ for low doping concentrations ($x$
$\leqslant$ 0.176) reveal that at a critical concentration, 0.102
$<$ $x$ $<$ 0.118, the tetragonal-to-orthorhombic transition
abruptly disappears whereas magnetic ordering with a propagation
vector of ($\frac{1}{2}$ $\frac{1}{2}$ 1) persists.  Among all of
the iron arsenides this observation is unique to Mn-doping, and
unexpected because all models for "stripe-like" antiferromagnetic
order anticipate an attendant orthorhombic distortion due to
magnetoelastic effects. We discuss these observations and their
consequences in terms of previous studies of
Ba(Fe$_{1-x}$$TM$$_x$)$_2$As$_2$ compounds ($TM$ = Transition
Metal), and models for magnetic ordering in the iron arsenide
compounds.
\end{abstract}

\maketitle

Recent systematic neutron and x-ray diffraction studies of
underdoped Ba(Fe$_{1-x}$Co$_x$)$_2$As$_2$ superconductors have
revealed fascinating results regarding the interactions among
structure, magnetism and superconductivity.  The undoped
\emph{AE}Fe$_2$As$_2$ parent compounds (\emph{AE} = Ba, Sr, Ca)
manifest simultaneous transitions from a high-temperature
paramagnetic tetragonal phase to a low-temperature orthorhombic
antiferromagnetic (AFM)
structure.\cite{Huang_2008,Jesche_2008,Goldman_2008} Upon doping
with Co for Fe in Ba(Fe$_{1-x}$Co$_x$)$_2$As$_2$,\cite{Sefat_2008}
both the structural (at $T_S$) and AFM ordering (at $T_N$) are
suppressed to lower temperatures and split, with $T_S$ slightly
higher than $T_N$.\cite{Ni_2008,Chu_2009,Lester_2009,Canfield_2010}
Neutron and x-ray studies have clearly established that both the
magnetic ordering and orthorhombic distortion are sensitive to
superconductivity throughout the Ba(Fe$_{1-x}$Co$_x$)$_2$As$_2$
series.\cite{Pratt_2009,Christianson_2009,Fernandes_2010,Nandi_2010}
At a given Co-composition, as the sample temperature is reduced
below the superconducting transition (at $T_c$), there is a clear
suppression of the magnetic order parameter, and reentrance into the
paramagnetic phase is observed for a Co-doping concentration of $x$
$\simeq$ 0.059.\cite{Fernandes_2010} Similarly, the magnitude of the
orthorhombic lattice distortion decreases below $T_c$ and reentrance
into the tetragonal structure was observed for $x$ $\simeq$
0.063.\cite{Nandi_2010} This striking behavior for
Ba(Fe$_{1-x}$Co$_x$)$_2$As$_2$ has been related to the strong
coupling between superconductivity and magnetism as well as an
unusual magnetoelastic coupling that arises from emergent nematic
order in the iron
arsenides\cite{Fang_2008,Xu_2008,Fernandes_2010_2}. The separation
of $T_S$ and $T_N$ and suppression of the magnetic order parameter
below $T_c$ have been confirmed for electron-doped
Ba(Fe$_{1-x}$Rh$_x$)$_2$As$_2$\cite{Kreyssig_2010} and
Ba(Fe$_{1-x}$Ni$_x$)$_2$As$_2$ as well.\cite{Wang_2010}

In strong contrast to what is found for the electron-doped
$AE$Fe$_2$As$_2$ compounds, hole doping on the Fe site through the
introduction of Cr\cite{Sefat_2009,Budko_2009} and
Mn\cite{Kim_2010,Liu_2010} has, so far, failed to produce
superconducting samples for any doping level, although
superconductivity is realized by hole-doping through the
substitution of K for the $AE$.\cite{Rotter_2008,Ni_2008a} This
indicates that the number of additional electrons (or holes) is not
the sole controlling factor for superconductivity. Furthermore,
unlike the suppression and eventual elimination of magnetic ordering
with increasing $x$ found for electron-doped compounds, recent
neutron studies of Ba(Fe$_{1-x}$Cr$_x$)$_2$As$_2$\cite{Marty_2010}
indicate that, for $x$ $\geqslant$ 0.30, the "stripe-like" AFM
structure is replaced by the G-type "checkerboard-type" structure as
found for BaMn$_2$As$_2$\cite{YSingh_2009} and proposed for
BaCr$_2$As$_2$.\cite{DSingh_2009}  Given the strong coupling between
structure, magnetism and superconductivity already established for
the iron arsenides, such differences in magnetic and structural
behavior in hole-doped materials demand attention.

Here we report on neutron and x-ray diffraction studies, together
with resistance measurements, of Ba(Fe$_{1-x}$Mn$_x$)$_2$As$_2$ for
low doping concentrations ($x$ $\leqslant$ 0.176).  We find that
within a critical concentration range, 0.102 $<$ $x$ $<$ 0.118, the
tetragonal-to-orthorhombic transition abruptly disappears while
magnetic ordering with a propagation vector of ($\frac{1}{2}$
$\frac{1}{2}$ 1) persists along with changes in the temperature
evolution of the AFM ordering.  The presence of "stripe-like" AFM
order in the absence of the orthorhombic distortion is
unanticipated, and holds important consequences for models of
magnetic ordering in the iron arsenides.

\begin{figure}
\begin{center}
\includegraphics[clip, width=.45\textwidth]{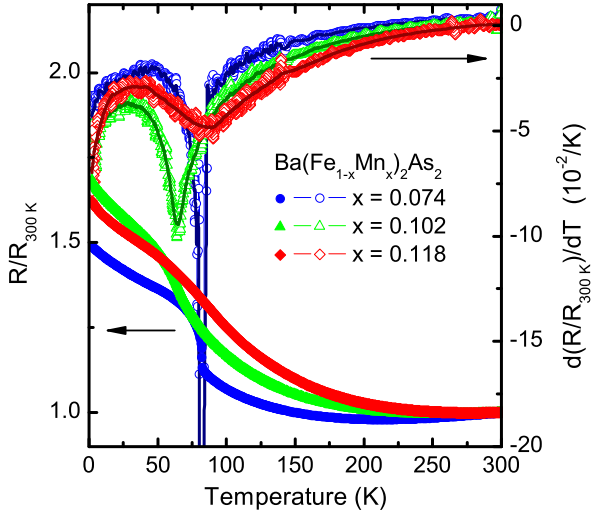}\\
\caption{(Color online) Resistance, normalized to the value at $T =
300$~K, and the temperature derivative of the resistance ratio for
the $x$ = 0.074, 0.102 and 0.118 samples.  Lines are guides to the
eye.} \label{bulk}
\end{center}
\end{figure}

Single crystals of Ba(Fe$_{1-x}$Mn$_x$)$_2$As$_2$ (0 $<$ $x$ $<$
0.176) were grown out of a FeAs self-flux using conventional
high-temperature solution growth.\cite{Ni_2008,Canfield_2010} Each
sample was measured at between 10 and 20 positions using wavelength
dispersive spectroscopy (WDS) to determine the Mn-doping
composition, $x$, with an uncertainty of 5$\%$. All samples used for
the neutron and x-ray measurements exhibited small mosaicities ($<$
0.02$^\circ$ full-width-at-half-maximum [FWHM]) measured by x-ray
rocking scans, demonstrating excellent sample quality.
Temperature-dependent AC electrical resistance data ($f$ = 16 Hz,
$I$ = 3 mA) were collected in a Quantum Design Magnetic Properties
Measurement System using a LR700 resistance bridge. In
Fig.~\ref{bulk} we show the resistance data (solid symbols)
normalized to their room temperature values, and the their
temperature derivatives (open symbols) for a representative subset
of three compositions, $x$ = 0.074, 0.102 and 0.118.  A sharp
anomaly, characteristic of all samples for $x$ $\leqslant$ 0.074 is
found at approximately 80~K for $x$ = 0.074, which broadens and
shifts to lower temperature for $x$ = 0.102 and then to higher
temperature for $x$ = 0.118.  If we associate these features with
magnetic and/or structural
transitions,\cite{Canfield_2010,Ni_2008,Pratt_2009} the
non-monotonic behavior of the characteristic temperature is highly
unusual for the iron arsenides. Only a single feature is observed in
the derivative curve indicating that the magnetic and structural
transitions are likely coincident in temperature, and
superconductivity is absent in all samples for $T$ $\geq$ 2~K.

\begin{figure}
\begin{center}
\includegraphics[clip, width=.45\textwidth]{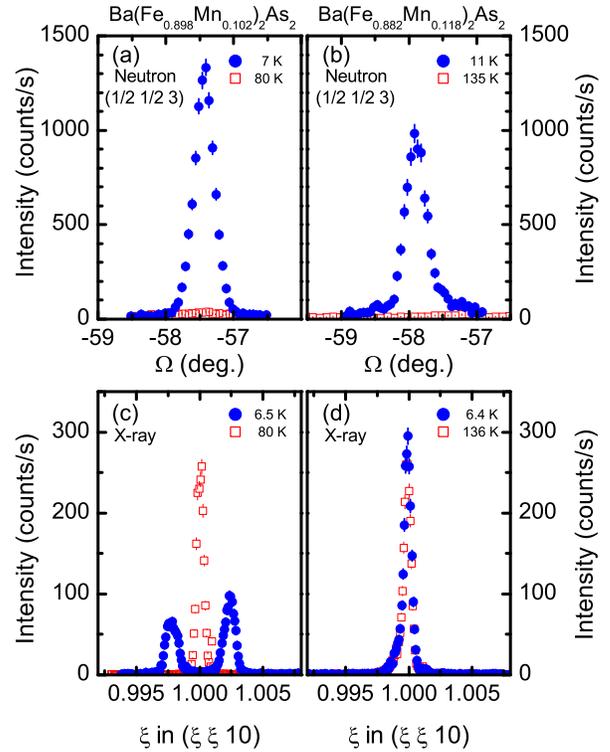}\\
\caption{ Neutron diffraction rocking scans through the ($1/2$ $1/2$
$3$) magnetic Bragg peak above (open squares) and below (filled
circles) the AFM transition for (a)
Ba(Fe$_{0.898}$Mn$_{0.102}$)$_2$As$_2$ and (b)
Ba(Fe$_{0.882}$Mn$_{0.118}$)$_2$As$_2$. Panels (c) and (d) show
scans along the ($\xi$, $\xi$, 0) direction through the (1 1 10)
charge reflection above (open squares) and below (filled circles)
the AFM transition for these samples. Note the splitting for the $x$
= 0.102 sample and its absence for $x$ = 0.118.} \label{data}
\end{center}
\end{figure}

High-resolution, single-crystal x-ray diffraction measurements were
performed on a four-circle diffractometer using Cu $K_{\alpha1}$
radiation from a rotating anode x-ray source, selected by a
germanium (1~1~1) monochromator. The diffraction data were obtained
between room temperature and 6~K, the base temperature of the
closed-cycle displex refrigerator. Neutron diffraction measurements
were performed on the HB1A diffractometer at the High Flux Isotope
Reactor at Oak Ridge National Laboratory. The experimental
configuration was 48'- 40'- 40'-136' with fixed incident neutron
energy of 14.7 meV, and two pyrolytic graphite filters for the
elimination of higher harmonics in the incident beam.

\begin{figure}
\begin{center}
\includegraphics[clip, width=.45\textwidth]{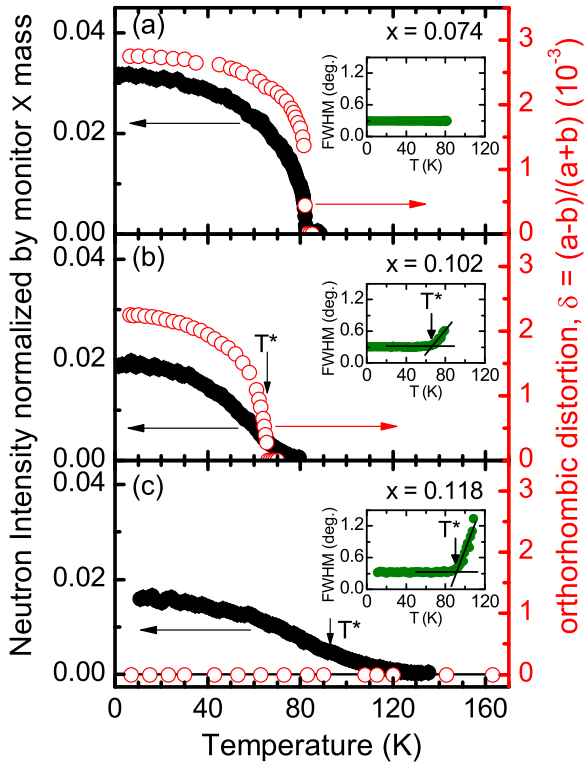}\\
\caption{(Color online) Temperature dependence of the integrated
intensities of the ($\frac{1}{2}$ $\frac{1}{2}$ 3) magnetic Bragg
peak (filled circles) and the orthorhombic distortion (open circles)
measured at the (1 1 10) charge peak positions for (a) $x$ = 0.074,
(b) $x$ = 0.102 and (c) $x$ = 0.118.  The insets to each panel show
the temperature dependence of the broadening of the ($\frac{1}{2}$
$\frac{1}{2}$ 3) magnetic peak and the definition of $T^*$.}
\label{ordpar}
\end{center}
\end{figure}

The principal results of our scattering studies are summarized in
Figs.~\ref{data} and \ref{ordpar} for a representative subset of the
compositions, $x$ = 0.074, 0.102 and 0.118. The neutron diffraction
data in Figs.~\ref{data}(a) and (b) show the magnetic Bragg peak at
($\frac{1}{2}$ $\frac{1}{2}$ 3) (using indices referenced to the
high-temperature tetragonal unit cell) for both $x$ = 0.102 and $x$
= 0.118, consistent with the "stripe-like" AFM order found for the
iron arsenide compounds. However, the x-ray data in
Figs.~\ref{data}(c) and (d) demonstrate that the orthorhombic
distortion, evident from the splitting of the (1 1 10) charge peak
for the $x$ = 0.102 composition, was not observed for $x$ = 0.118.
Figure~\ref{ordpar} displays the temperature evolution of the
magnetic order, measured by neutron diffraction, and the
orthorhombic distortion, measured by x-ray diffraction, for these
same compositions.  The integrated intensity of the magnetic
scattering (filled circles) was measured at the ($\frac{1}{2}$
$\frac{1}{2}$ 3) magnetic Bragg position as the sample angle was
scanned [see Figs.~\ref{data}(a) and (b)].  The orthorhombic
distortion, $\delta$, was calculated from the splitting of peaks
observed in ($\xi$ $\xi$ 0)-scans through the (1 1 10) Bragg peak
[see Fig.~\ref{data}(c) and (d)]. For samples with $x$ $\leqslant$
0.074 [Fig.~\ref{ordpar}(a)], we observe well defined AFM and
structural transitions that are, within our resolution, coincident
in temperature.  For $x$ = 0.102 [Fig.~\ref{ordpar}(b)], a weak
"tail" of magnetic scattering extends to temperatures above the
structural transition and, for $x$ $\geqslant$ 0.118, the structural
transition is absent (the sample remains tetragonal down to at least
$T = 6.4~K$ within our resolution for $\delta$ of 1 $\times$
10$^{-4}$) and the temperature evolution of the AFM order is quite
different from what is observed for $x$ = 0.074. For $x$ $\geqslant$
0.118, a distinct broadening of the magnetic peak beyond the
resolution of our measurement is observed for temperatures above
$T^*$, as defined below and in the insets to Figs.~\ref{ordpar}(b)
and (c).

\begin{figure}
\begin{center}
\includegraphics[clip, width=.45\textwidth]{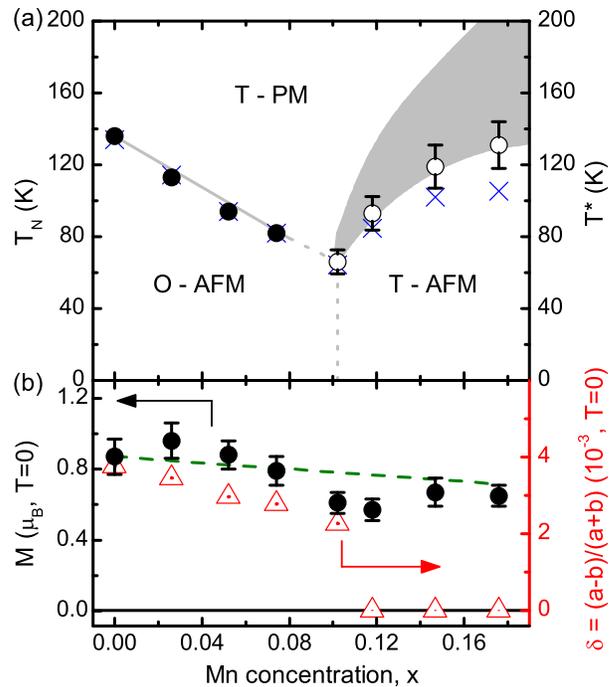}\\
\caption{(Color online) (a) The compositional phase diagram for
Ba(Fe$_{1-x}$Mn$_x$)$_2$As$_2$ determined from neutron and x-ray
diffraction measurements.  Closed circles denote $T_N$ and open
circles represent $T^*$ as described in the text.  Crosses denote
the temperature corresponding to minima of $\frac{dR}{dT}$ found in
Fig.~\ref{bulk}.  The shaded region denotes the extent of the
magnetic scattering above $T^*$. The vertical dashed line marks the
approximate composition for the change from an orthorhombic to
tetragonal structure. (b)The measured magnetic moment and structural
distortion as a function of Mn-doping.  The dashed line represents
the value of the magnetic moment per Fe atom rather than Fe/Mn site
as a function of Mn-doping.} \label{PD}
\end{center}
\end{figure}

In Fig.~\ref{PD}(a) we have used the neutron, x-ray and resistance
data to construct a phase diagram in the low Mn-doping regime for
Ba(Fe$_{1-x}$Mn$_x$)$_2$As$_2$.  The phase line between the
paramagnetic/tetragonal and AFM/orthorhombic phase for $x$
$\leqslant$ 0.074 was easily determined from the well-defined onset
of the distortion and the appearance of a resolution limited
magnetic Bragg peak at ($\frac{1}{2}$ $\frac{1}{2}$ 3).  For $x$
$\geqslant$ 0.102, however, the onset of long-range magnetic order
is more difficult to identify.  Therefore, we have defined a
characteristic temperature, $T^*$, which denotes the temperature
below which the width of the magnetic peak is limited by our
instrumental resolution (approximately 0.3$^{\circ}$ FWHM). We note
that the values of $T^*$ follow the same trend seen for the maxima
in $\frac{dR}{dT}$ in Fig.~\ref{bulk}. The gray band in the phase
diagram represents the temperature range, above $T^*$, where
magnetic scattering at ($\frac{1}{2}$ $\frac{1}{2}$ 3) persists [See
Figs.~\ref{ordpar}(b) and (c)].

In Fig.~\ref{PD}(b) we plot the measured structural distortion and
the magnetic moment per Fe/Mn site, extrapolated to $T$ = 0 as
described in our previous work,\cite{Fernandes_2010} as a function
of doping concentration. Several interesting comparisons can be made
between these results and previous x-ray and neutron scattering
studies of
Ba(Fe$_{1-x}$Co$_x$)$_2$As$_2$.\cite{Lester_2009,Pratt_2009,Christianson_2009,Fernandes_2010,Nandi_2010}
First, we note that our data for Ba(Fe$_{1-x}$Mn$_x$)$_2$As$_2$ for
$x$ $\leqslant$ 0.074 unambiguously show that the structural and
magnetic transitions remain locked together, unlike the separation
of the structural and AFM transitions found for Co-doping.
Furthermore, at $x$ = 0.102, we find a broadened magnetic peak at
($\frac{1}{2}$ $\frac{1}{2}$ 3) \emph{above} the structural
transition and, for $x$ $\geqslant$ 0.118, we observe the magnetic
Bragg peak at ($\frac{1}{2}$ $\frac{1}{2}$ 3) in the absence of an
orthorhombic distortion, a surprising observation that will be
discussed below. Finally we note that the magnetic moment per Fe/Mn
site as well as the magnitude of the structural distortion vary only
weakly with composition for $x$ $\leqslant$ 0.102 whereas, for
Co-substitution, the suppression of the magnetic moment and
structural distortion with doping is much more severe.

It is also useful to compare these results to what has recently been
found for Ba(Fe$_{1-x}$Cr$_x$)$_2$As$_2$.\cite{Marty_2010}  At much
higher Cr concentrations, $x$ $\geqslant$ 0.30, Ref.
~\onlinecite{Marty_2010} reports that the "stripe-like" magnetic
structure is replaced by G-type, "checkerboard," magnetic order as
shown by polarized and unpolarized neutron diffraction measurements
of the integrated intensity of the (1 0 1) Bragg peak (Fig. 3 in
Ref. \onlinecite{Marty_2010}).  G-type AFM order has been proposed
for the parent BaCr$_2$As$_2$ compound,\cite{DSingh_2009} and
measured for BaMn$_2$As$_2$,\cite{YSingh_2009} so it is not
unreasonable to expect this change in magnetic structure at high
enough Cr-, or Mn-, doping.  However, our unpolarized neutron
diffraction measurements of the (1 0 1) peaks for the highest Mn
concentrations, $x$ = 0.147 and $x$ = 0.176, find no evidence of
G-type ordering below $T = 300$~K.  More specifically, we find no
significant change in the (1 0 1) peak between 12~K and 300~K. We
can not exclude G-type ordering that develops well above room
temperature given the high ordering temperature of the parent
compound,\cite{YSingh_2009} but view this as unlikely in light of
the substantial dilution of Mn in our samples. For both Cr- and
Mn-doping, the moment per Fe-site remains constant (Cr), or
decreases only weakly (Mn) with increasing concentration up to $x$
$\approx$ 0.20. Indeed, as the dashed line in Fig.~\ref{PD}(b)
shows, the decrease in the measured moment is consistent with the
decreasing Fe concentration implying that the Mn moment does not
contribute to the magnetic AFM order characterized by the
($\frac{1}{2}$ $\frac{1}{2}$ 1) propagation vector. Furthermore, for
Mn-doping we find an increase in the characteristic temperature
($T^*$) associated with magnetic ordering with this propagation
vector for $x$ $>$ 0.102, whereas for Cr-doping, the ordering
temperature for this propagation vector continues to decrease until
the transition is completely suppressed at $x$ = 0.335 where the
G-type AFM structure is observed.\cite{Marty_2010} All of this
points to interesting differences in the phase diagrams between
Ba(Fe$_{1-x}$Mn$_x$)$_2$As$_2$ and Ba(Fe$_{1-x}$Cr$_x$)$_2$As$_2$.

The observation of a magnetic structure characterized by a
propagation vector of ($\frac{1}{2}$ $\frac{1}{2}$ 1) in the absence
of an orthorhombic distortion (for $x$ $>$ 0.102) is very surprising
and unique to Ba(Fe$_{1-x}$Mn$_x$)$_2$As$_2$ among the iron
arsenides; models for "stripe-like" AFM order in the iron arsenides
anticipate an attendant orthorhombic distortion due to
magnetoelastic effects.\cite{Fang_2008,Xu_2008,Nandi_2010}
Furthermore, this observation is difficult to reconcile with current
theories that promote orbital ordering\cite{Lv_2009,Chen_2010} as
the driving force for the "stripe-like" magnetic phase and the
orthorhombic distortion. A second key result of this study is the
qualitative change in the temperature dependence of the magnetic
ordering for compositions in excess of $x$ = 0.102 and the distinct
broadening of the magnetic peak for $T > T^*$.  At this point it is
not clear whether the scattering above $T^*$ for $x$ $>$ 0.102 is
purely elastic or has a quasielastic component within the finite
energy window of our neutron measurements, a point that should be
investigated further.

The change in the temperature dependence of the magnetic peak points
to a strong perturbation of the magnetic ordering, perhaps through
disorder effects associated with the introduction of the more
localized Mn moments. Furthermore, the abruptness of this change
with composition (over a narrow range of $\Delta x$ $<$ 2$\%$)
offers the intriguing possibility that the magnetic structure of
Ba(Fe$_{1-x}$Mn$_x$)$_2$As$_2$ is modified for $x$ $>$ 0.102. In
recent theoretical work, Eremin and Chubukov\cite{EandC_2010} point
out that a generic spin configuration for the magnetic iron layers
has the form,
${\mathbf{\Delta}_\mathrm{{1}}\mathrm{e^{i\mathbf{Q}_\mathrm{{1}}\cdot
\mathbf{R}}} +
\mathbf{\Delta}_\mathrm{{2}}\mathrm{e^{i\mathbf{Q}_\mathrm{{2}}\cdot
\mathbf{R}}}}$, where $\mathbf{\Delta}_\mathrm{{1}}$ and
$\mathbf{\Delta}_\mathrm{{2}}$ correspond to two order parameters
for ordering at wavevectors $\mathbf{Q}_\mathrm{{1}} = (0,\pi)$ and
$\mathbf{Q}_\mathrm{{2}} = (\pi,0)$, respectively, in the unfolded
Brillouin zone.  The observed "stripe-like" magnetic structure
occurs when ${\mathbf{\Delta}_\mathrm{{1}}}$ = 0 and
${\mathbf{\Delta}_\mathrm{{2}}}~\parallel~\mathbf{Q}_\mathrm{{2}}$.
However, when they consider a coupling between the second hole
pocket at the $\Gamma$ point with the elliptical electron pocket at
$(0,\pi)$, a two-\textbf{Q} structure with both
$\mathbf{\Delta}_\mathrm{{1}}~\neq~0$ and
$\mathbf{\Delta}_\mathrm{{2}}~\neq~0$ can emerge. For
$\mathbf{\Delta}_\mathrm{{1}}~\perp ~\mathbf{\Delta}_\mathrm{{2}}$
and
$|{\mathbf{\Delta}_\mathrm{{1}}}|~=~|{\mathbf{\Delta}_\mathrm{{2}}}|$,
this two-\textbf{Q} structure does not break the tetragonal symmetry
and, therefore, does not yield an orthorhombic distortion of the
lattice, consistent with our results. Because of the presence of
magnetic domains in the tetragonal phase, magnetic peaks for the
"stripe-like" and two-\textbf{Q} AFM structures can not be
distinguished.

We acknowledge valuable discussions with J. Schmalian and R. M.
Fernandes.  This work was supported by the Division of Materials
Sciences and Engineering, Office of Basic Energy Sciences, U.S.
Department of Energy. Ames Laboratory is operated for the U.S.
Department of Energy by Iowa State University under Contract No.
DE-AC02-07CH11358. The work at the High Flux Isotope Reactor, Oak
Ridge National Laboratory (ORNL), was sponsored by the Scientific
User Facilities Division, Office of Basic Energy Sciences, U.S.
Department of Energy (U.S. DOE). ORNL is operated by UT-Battelle,
LLC for the U.S. DOE under Contract No. DE-AC05-00OR22725.

\bibliographystyle{apsrev}
\bibliography{bamnfe2as2_arxiv}

\end{document}